# MVC-3D: Adaptive Design Pattern for Virtual and Augmented Reality Systems

Samir Benbelkacem, Djamel Aouam, Nadia Zenati-Henda
Centre de Développement des technologies Avancées.
Cité 20 août 1956, Baba-Hassen, Algiers
sbenbelkacem@cdta.dz

Abdelkader Bellarbi, Ahmed Bouhena
Centre de Développement des technologies Avancées.
Cité 20 août 1956, Baba-Hassen, Algiers
abellarbi@cdta.dz

Samir Otmane
IBISC, Univ Evry, Université Paris-Saclay
91025, Evry, France
samir.otmane@ibisc.univ-evry.fr

## Abstract

In this paper, we present MVC-3D design pattern to develop virtual and augmented (or mixed) reality interfaces that use new types of sensors, modalities and implement specific algorithms and simulation models. The proposed pattern represents the extension of classic MVC pattern by enriching the View component (interactive View) and adding a specific component (Library). The results obtained on the development of augmented reality interfaces showed that the complexity of M, iV and C components is reduced. The complexity increases only on the Library component (L). This helps the programmers to well structure their models even if the interface complexity increases. The proposed design pattern is also used in a design process called MVC-3D in the loop that enables a seamless evolution from initial prototype to the final system.

## 1   Introduction

The evolution of the interaction between a user and mechatronic systems require a new design of task-adequate and usable user interfaces. This leads to a new generation of user interfaces (NGUIs) that provide a set of new technologies of human-computer interaction based on newly developed modalities, sensors and algorithms. NGUIs diverge from the well-known WIMP (Window Icon, Menu and Pointer) and use novel interaction paradigms such as Virtual Reality (VR) and Augmented Reality (AR), tangible, embodied and multi-modal interfaces. In recent years, researches focus on technical issues such as, tracking, rendering, I/O devices in order to design such interfaces. The lack of design models and interdisciplinary nature of virtual and augmented (or mixed) reality design require design approaches based on iterative refinement. Also, the lack of formal verification techniques and the limited design expertise limits the applicability of techniques like experts' reviews for NGUI applications, except for experimental evaluation through tests with end-users that could be a potential solution. However, a simple implement and test approach is not viable because the implementation of working prototype is expensive and time consuming, limiting the number of concepts and designs that can be explored. To overcome these problems, we propose an adaptive design pattern, with supplementary concepts specific to virtual and augmented reality systems. This paper briefly illustrates the principal idea of our design approach in section 3 and presents preliminary results in section 4. But we start with a short review of related work in section 2.

## 2   Related work

Design models have been proposed for the development of virtual and augmented reality interfaces in the literature [Emm 17]. Ishii [Ish 08] extended the MVC model [Bur 92] to develop tangible interfaces. Stöcklein [Sto 09] proposed MVCE model that integrates "Environment" component into the MVC model. In [Tar 97], AMF-C design pattern was proposed to design mixed reality collaborative systems. In our previous works [Ben 14], we realized three implementations of a fragment of the same application "car engine maintenance assistance using Augmented Reality". The first implementation used a traditional MVC model [Bur 92]. The second implementation used an MVC pattern implemented using a communication pattern proposed by Eckstein [Eck 07]. The third implementation was based on an AMF-C architecture model [Tar 97]. The programmers estimated that structuring the code into components can be complex using the MVC model. The complexity of components increases when implementing AMF-C model. In the other hand, we observed that the MVCE model [Sto 09] increases the complexity of components when the interface contains more 3D information and integrates many processing (e.g. simulation tools, tracking and gesture recognition algorithms). In addition, introducing complex simulation models and heavy calculations in Model (M) and Environment (E) components makes the development tedious and upsets programmers' practices.

In our approach, we integrate an additional component to MVC model that consider all tools, SDK and heavy processing separately. When a technology is modified (simulation toolkits, algorithms and devices), we change only the additional component's content without affecting the Controller and the Model. Only adaptation can be made for these two components to support the changes. With this approach, we can reduce the complexity and recurrent accesses to MVC components and preserve the programmers' practices.

# 1 Design Approach

The goal of many augmented and virtual-reality applications is the development of better user interfaces based on an iterative design process. However, most design processes assume that the underlying technology is well defined and stable, a condition that is often not respected in the development of NGUIs which use developing technologies that are still in early experimental stages. Changes in the underlying technologies can be problematic, if a design process does not anticipate such volatility and therefore provides no means to handle it in a systematic and structured way. Also, most of the virtual and augmented reality prototypes are designed without any structuration and using a Trial-and-Error prototyping approach. This leads to an overhead in the programming, since parts of the application cannot be reused and must be re-implemented for each prototype. In our approach, we have developed an iterative design process which can be used for NGUIs based on our proposed MVC improved model.

1.1 MVC-3D Design Pattern

We present an adaptive design pattern to implement interaction techniques of virtual and augmented reality interfaces. We propose a structure that extends the (MVC) model-view-controller design pattern with additional components and features that consider independently all processing specific to VR & AR (e.g. tracking algorithms) and integrate the environment (virtual and real) with their corresponding interactive devices. For tangible interfaces we are based on the concepts presented in [Ish 08].

The benefit of the classical MVC pattern is the separation of interaction and visualization aspects of the user interface when designing the application. Also, it enables modular designs with which changes performed on a component doesn't affect other components. The Model (M) encapsulates the application data and the functionalities of the application. The Controller (C) handles the user actions and plays an intermediate role between View and Model. The View (V) represents the visual items of the application. One of the key elements of virtual and augmented (or mixed) reality user interfaces is the maintaining of coherent relationship between real world and virtual models in real time when incorporating heterogeneous software and hardware technologies. To guarantee this coherence, the structure of the components should be less complex as possible to facilitate the information exchange and reduce processing time. Our vision is to integrate heavy processing of the application in a specific component which could be modular and reusable. Therefore, we introduced an additional component "Library" to MVC model (see Fig.1) that encapsulates all processing using specific algorithms (e.g. tracking techniques, gestures, faces and speech recognition algorithms), complex simulation models (e.g. Matlab/Simulink) and SDK/toolkits to process the amount of data provided by the View via the Controller.

In addition, some interaction devices should be in the user's view field to perform 3D manipulation tasks (gestures interaction, speech control…). From practical point of view, we observed that it is difficult to dissociate physically both the environment and interactive devices from the view of the user. For that purpose, we brainstormed to conceptually merge these elements. We proposed to enhance the View component by introducing a sub-component that captures the real environment model of the application and a sub-component that integrates the sensors module and manages the tangible objects (physical and digital). We obtained then a new component that we call interactive View (iV) (see Fig.1).

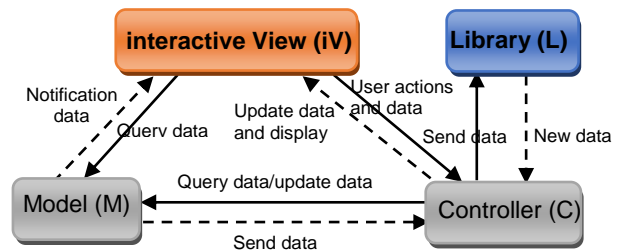

Figure1: MVC-3D Model

Using the MVC-3D structure, components can be refined independently. One key benefit is the possibility to develop an interactive user interface along the mixed reality continuum [Mil 94] in which dedicated processing is formulated by a component: Library (L). This component involves heterogenous and complexe simulation models, SDKs and algorithms. The benefit of this component is the capability to manage this heterogeneity and the relationship between the different models. If we plan to change an algorithm or a toolkit, we just modify the Library content without completely changing the Model and Controller's content. A simple adaptation could be made.

1.2 Design Process

Our goal is to iteratively develop virtual and augmented reality prototypes that could be refined to correspond the user's expectations. These prototypes can be then used to evaluate the system with end-users and to validate the technologies used in the system. Using the MVC-3D model, components can be refined independently. The benefit of this approach is the possibility to develop a user interface along the mixed reality continuum [Mil 94] starting from purely virtual environment to real one. Figure 2 describes our design process.

We define each MVC-3D component as an 'entity'. An entity can be anything from a model, a visual representation, to a controller.

At the beginning of the development process an initial set of entities is identified. For each entity in this set the inputs and outputs are defined. If the development targets a complex mechatronic system, each relevant system component (either hardware or software) is initially represented by an entity. Additional entities represent the elements of the user interface. This initial set of entities can later be extended. It should include all entities needed for a first prototype that provides a rough approximation of the planned system. The first prototype is then composed from these entities, connecting the information flow between the input and output ports as required by the application. For the technical implementation we use our mixed reality environment described in [Ben 14].

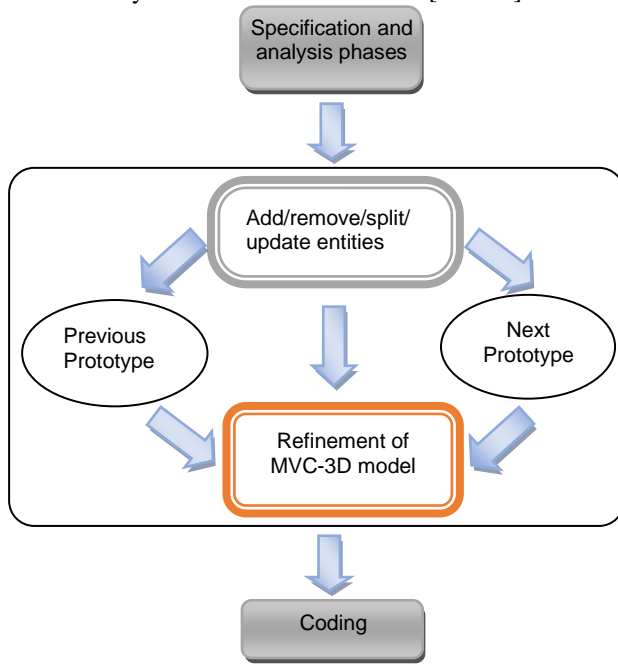

Figure2: MVC-3D in the loop

Typically, the first prototype consists of an environment in which all entities are implemented purely in software.

The resulting system would therefore be a purely virtual environment. The key benefit is that elements in a virtual environment are faster and cheaper to develop. Depending on the development goals and priorities, entities are selected for refinement.

Refinement means that either the behavior or visual representation of an entity is updated (e.g. the 3D-model) or an entity replaced by another version of the same entity (e.g. Game physics vs. Matlab/Simulink). If the entity is concerned with simulating real-world elements, typical refinements would be the replacement of a simple simulation with a more realistic one, or the replacement of a simulation component with its real world equal. This approach allows to move from purely virtual environments to hardware in the loop (or more precisely mixed reality in the loop) systems in a structured way. If an entity is concerned with the implementation of interaction or visualization techniques the replacement could either be a complete exchange of the component (e.g. to compare alternative approaches to system control) or stepwise refinements, in which a user interface is refined according to user feedback as in established iterative user centered design processes. As the development progresses from an initial basic prototype to more complex systems, it can also be necessary to adjust the number of entities, e.g. by splitting the functionality of an entity into two or more, or by adding or removing entities. In all cases the data flow connections between the entities must be checked and corrected consequently.

## 2 Preliminary Results

We have applied the MVC-3D model during the implementation phase of the extended 2TUP process [Ben 14] for the fragment of the application "maintenance of a car engine by AR". We have developed three prototypes with different interaction techniques using MVC-3D pattern for each one. Preliminary results are given in Table 1.

Prototype 1 shows an AR interface of a real cylinder head of a car engine on which are displayed four 3D pistons (interactive View (iV)). The Model (M) encapsulates 3D pistons' Data, the transformation matrix of the 3D pistons and the markers' Data. The Controller (C) manages the images' data provided by the camera, sends the data to the Library (L) and receives the results. The Library (L) uses ARtoolKit tracking algorithm to process the data provided by the Controller (C).

Prototype 2 displays the same augmented scene, but without markers (interactive View (iV)). The Library (L) implement MOBIL algorithm [Bel 14],[Bel 17] for marker-less tracking system. The programmers add to the Model (M), already established for prototype 1, the reference image' data of the cylinder head. The Controller (C) was adjusted to support exchanging data with the Library supporting MOBIL algorithm. In this case, the content of the Library was modified by implementing MOBIL algorithm. Therefore, the complexity increases only at the Library (L) component in which processing to detect and track the scene is more consistent. Prototype 2 is the refinement of prototype 1 using the design process given in Fig.2. In fact, we have removed an entity (ARtoolKit SDK) and replaced it by a new entity (MOBIL Marker-less technique). So, the MVC-3D is refined along Library (L) axis (see row 2, column 4 of Table 1).

Prototype 3 is an extension of prototype 2. It provides an interactive interface where the user manipulates the 3D

Table1: Implementation of three prototypes using MVC-3D model

| Model (M) | Controller (C) | interactive View (iV) | Library (L) | Complexity measure | Prototype |
|---|---|---|---|---|---|
| - Transformation matrix of the 3D Pistons.<br>- 3D pistons' Data<br>- Markers' Data. | - Processing image provided by camera: convert image (pre-processing) and inserts 3D pistons into image (post-processing). | - Four 3D pistons.<br>- Environment: cylinder head and square markers.<br>- Sensors module: camera. | - Recognition of simple square markers using ARtoolkit. | (complexity diagram) | (prototype image) |
| - Transformation matrix of the 3D Pistons.<br>- 3D pistons' Data.<br>- Reference image data (Cylinder head's image). | - Processing image provided by camera: convert image (pre-processing) and inserts 3D pistons into image (post-processing). | - Four 3D pistons.<br>- Environment: cylinder head.<br>- Sensors module: camera. | - Detection and recognition of reference image (cylinder head's image) using MOBIL algorithm.<br>- Camera pose estimation. | (complexity diagram) | (prototype image) |
| - Transformation matrix of the 3D Pistons.<br>- 3D pistons' Data.<br>- Reference image data (Cylinder head's image).<br>- Gestures' Data. | - Processing image provided by camera: convert image (pre-processing) and inserts 3D pistons into image (post-processing).<br>- Gesture based control. | - Four 3D pistons.<br>- Environment: cylinder head.<br>- Sensors module: camera and Kinect. | - Detection and recognition of reference image (cylinder head's image) using MOBIL algorithm.<br>- Camera pose estimation.<br>- Gesture recognition. | (complexity diagram) | (prototype image) |

pistons using his hands. The Library encapsulates, besides MOBIL tracking system, algorithms calculating the position and orientation of a hand and its tracking in the AR scene. The processing linked to the gesture recognition [Bel 13] was not integrated into the Controller (C) and the classical View (V), neither in the Model (M), but in the Library (L). The Model (M) is enriched by adding the gestures' data. The Controller (C) manages both the images and the gestures data which are provided by the Library (L). The interactive View (iV) encapsulates the Kinect module and the user's hands data besides of the real cylinder head's reference image data. Therefore, interactive View (iV) component facilitates data access (e.g. the Kinect delivers directly gestures' data to the classical View (V)). The Controller (C) can easily handle data arising from the real environment.

In the same manner of the precedent case, prototype 3 is the refinement of prototype 2. We have added to prototype 2 two entities: (1) a Kinect with 3Gear library that capture the gesture motion and (2) the AR - Head Mounted Display (HMD) to view the 3D piston manipulated by the user. So, the MVC-3D is refined along Model (M), interactive View (iV), Controller (C) and Library (L) axes (see row 3 of Table 1).

Comparing prototypes 2 and 3 using our approach, the structure and content of the Model and Controller have not deeply changed; a simple adaptation and/or enhancement have been made. Complex algorithms and models are integrated into the Library instead of Model and Controller comparing to the approaches presented in the literature where all the components increase in complexity. In the other hand, the structure of the interactive View can be relevant since it promotes exchanging data between the classical View, the environment model and interactive devices.

## 3 Conclusion

To consider virtual and augmented reality systems' specifications in a design process, a design approach is required. In this paper we presented the "virtual and augmented reality in the loop" process, based on structuring the VR & AR applications into Model, interactive View, Controller and Library components that can be refined individually. We have shown how this model was used successfully to refine prototypes. In the other hand, we detailed MVC-3D design model. This model can help programmers to better structure the programming process and focuses specific processing in the Library component with which the implementation of methods and tools can be well organized. MVC-3D can be also adapted for applications involving heterogeneous algorithms and different interaction devises. Using our design approach, we can preserve the programmers' practices and reduce programming complexity.